\documentclass[12pt]{iopart}
\usepackage{graphicx}
\def\rightmark{$J/\psi$ and $\Upsilon$ measurements in STAR}
\def\leftmark{M. R. Cosentino (STAR Collaboration)}

\begin{document} 
 
\title{$J/\psi$ and $\Upsilon$ measurements in STAR} 

\author{\footnotesize MAURO R. COSENTINO for the STAR COLLABORATION}

\address{Instituto de F\'{i}sica, Universidade de S\~ao Paulo, Rua do Mat\~ao - travessa R, 187 \\  S\~ao Paulo, SP 05508-090,Brazil}

\ead{mcosent@dfn.if.usp.br}

\pacs{13.20.Gd, 13.85.Ni, 14.40.Lb, 14.40.Nd, 25.75.Dw}

\hspace{1.7cm}{\it \footnotesize  Proceedings for the 24th Winter Workshop on Nuclear Dynamics}

\begin{abstract}
{Heavy-quarkonium states are expected to evidenciate the deconfinement of the nuclear matter into a Quark-Gluon Plasma in heavy-ion collisions. To strive  conclusive information from quarkonium production modification in A+A collisions, systematic measurements of the $J/\psi$ and $\Upsilon$ states in p+p, d+Au and Au+Au collisions are necessary. To accomplish this mission the STAR experiment has a Quarkonium program based on the development of specific trigger setups that take advantage of the large STAR acceptance. In this work we present the preliminary results of the $J/\psi$ and $\Upsilon$ measurement in 200 GeV p+p and the first measurements of $\Upsilon$ in 200 GeV heavy ion collisions.}
\end{abstract}

\setcounter{page}{1}

\section{Introduction}\label{intro}

Heavy Quarkonium states are of special interest in heavy ion collisions since it was noted by Matsui and Satz ~\cite{MatsuiSatz} that, in Quark-Gluon Plasma, the $J/\psi$ would dissociate by the effect of color screening, leading to a suppression of its production when compared to $J/\psi$ production in p+p collisions. More than twenty years have elapsed since then and many theoretical developments have been made in order to understand the heavy quarkonium states production in QGP and relativistic heavy-ion collisions.

One of these developments predict that there should be a production modification pattern of quarkonium states due to sequential suppression of the various bound $Q\bar{Q}$ states~\cite{SatzSeq}. In this scenario the excited quarkonium states would melt sequentially due to the different radii of the different states, leaving a sequential suppression pattern on the inclusive $J/\psi$ measurement.

Another feature recently proposed is that of quarkonium states produced at high momentum with respect to the QGP. In this scenario the quarkonium states are depicted at rest, and the medium moving in respect to it as hot wind~\cite{hotWind}. This study indicates that the screening radius $L_S$ decreases as $\gamma$ increases at a fixed temperature, allowing quarkonium states to melt at lower QGP temperatures than those predicted by the static quarkonium calculations.

In order to address the question about the quarkonium production modification in relativistic A+A collisions it is necessary to understand the baseline elementary production and the cold nuclear matter (CNM) effects. To achieve this knowledge the STAR collaboration developed a Quarkonium Program aimed to a systematic study of quarkonium production in p+p, d+Au, Au+Au and Cu+Cu collisions. One of the most important developed features of this Program is the specific trigger for heavy quarkonium states, detailed in section \ref{trigger}. 

In section \ref{results} we present the results of $J/\psi$ measurements in p+p and Cu+Cu, including an extended $p_T$ spectrum and the $\Upsilon$ (1S+2S+3S) production in p+p and Au+Au, the first in heavy ion collisions.
 
\section{Experimental Setup}\label{setup}  

STAR detector~\cite{star} is a multipurpose experiment composed of several subsystems suited to measure many different observables in the central rapidity region. Its large acceptance ($\vert\eta\vert<$1 and 0$<\phi<$2$\pi$)  assures STAR the capability of measure many of the quarkonium states, including the $J/\psi$ and $\Upsilon$ family, through their $e^+e^-$ decay channel. The subsystems used to do it are the Time Projection Chamber (TPC), the Barrel Electromagnetic Calorimeter (BEMC) and the Central Trigger Barrel (CTB). Also, a specific trigger was designed for $J/\psi$ and $\Upsilon$ measurements, while the high-$p_T$ $J/\psi$ measurement took advantage of pre-existing High Tower triggers.

The TPC is the main STAR subsystem. It is designed to reconstruct the tracks of charged particles, giving precise information on momentum and ionization energy loss ($dE/dx$). With the TPC it is possible to make a first set of particle identification cuts for electrons.

The BEMC is a sample calorimeter with full azimutal coverage, surrounding the TPC, and is divided in 120 modules of 40 towers each (20 in $\eta$, 2 in $\phi$.). Each module has  a range of -1$<\eta<$1 and have azimutal coverage of 2$\pi$/60. The towers are stacks of 21 alternating layers of scintilator and lead absorber, corresponding to $\sim$21$X_0$ and with geometrical acceptance of $\Delta\eta\times\Delta\phi=$0.05$\times$0.05 each. The energy resolution in the towers is $\frac{dE}{E}\sim\frac{16\%}{\sqrt{E}}$. The first 2 lead-scintilator layers compose the Pre Shower Detector (PSD) and at the position of 5$X_0$ from the beamline, sits the Shower Maximum Detector (SMD), which is a multi-wired proportional chamber detector with 2 planes of wires (in $\eta$ and $\phi$), providing good spacial resolution and the determination of the electromagnetic showers.

Finally, the CTB is a detector made of 120 trays containing 2 scintilator slats each. It is located between the TPC and the BEMC, with the same acceptance coverage of the last one. The CTB was originally developed to trigger events based on the detection of charged particles passing through it, but in this work it was used as an auxiliary system to trigger $J/\psi$ events. Each one of the scintilator slats matches the same area of $\sim$20 BEMC towers.

\subsection{Trigger}\label{trigger}

In order to accomplish the measurement of a rare probe such as $J/\psi$ and $\Upsilon$ we needed to develop a specific event trigger setup. This trigger setup was divided in two levels, the first of them (L0) was a topologic trigger designed in the hardware level. The details of this specific trigger are discussed in more detail in \cite{upsilon,mrcQM08}.

For the high-$p_T$ $J/\psi$ measurement the trigger setups used were simpler~\cite{zebo}: high tower triggers requiring that at least one BEMC tower have a reading above the threshold per event. For p+p collisions two thresholds were used, 3.5 and 5.4 GeV and for the Cu+Cu data the high tower threshold was of 3.75 GeV.

  \begin{figure}
    \begin{center}
      {
      \label{fig:jpsiLineShape}
      \includegraphics[width=0.48\textwidth]{{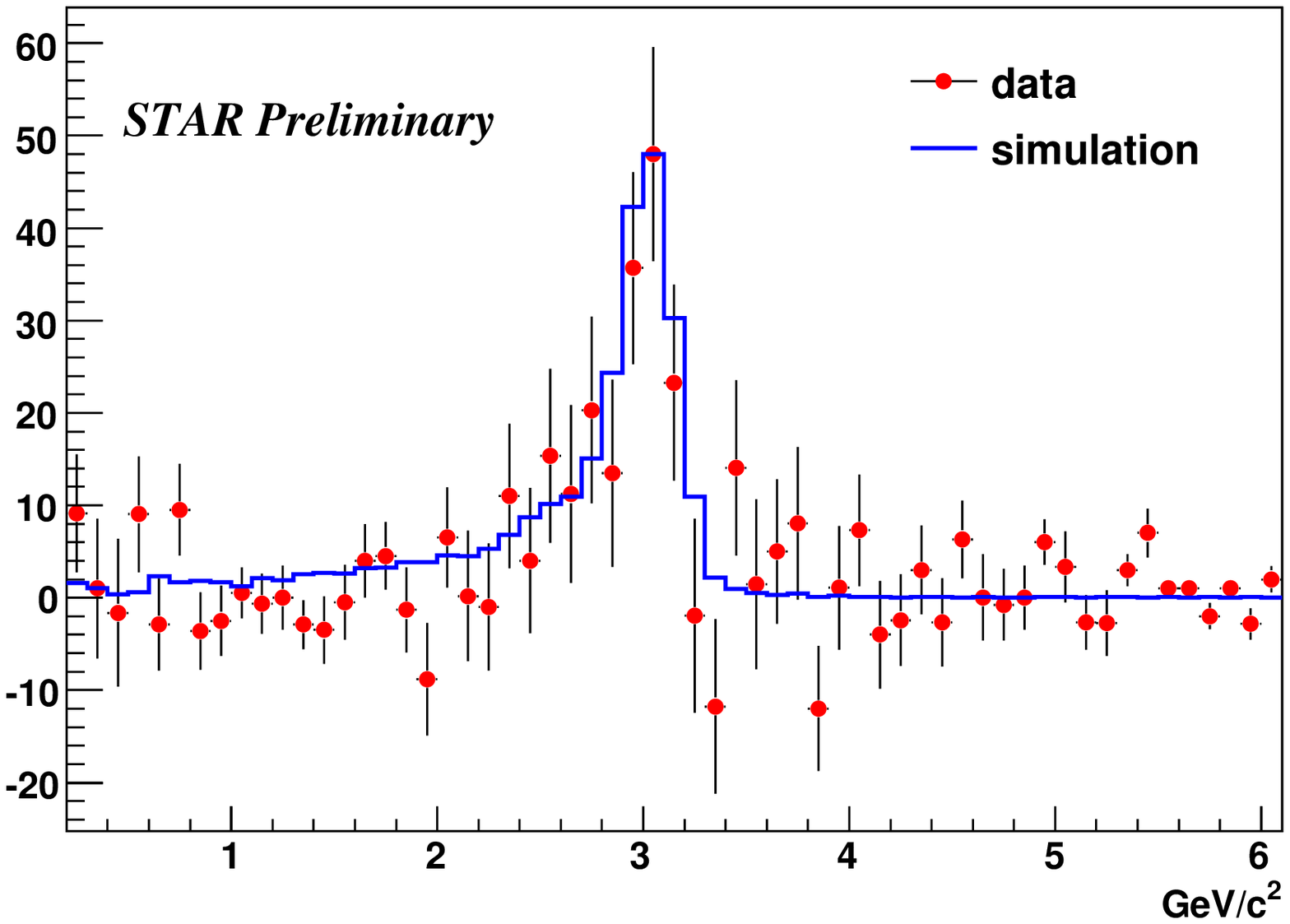}}
      }
      {
      \label{fig:jpsiPt}
      \includegraphics[width=0.48\textwidth]{{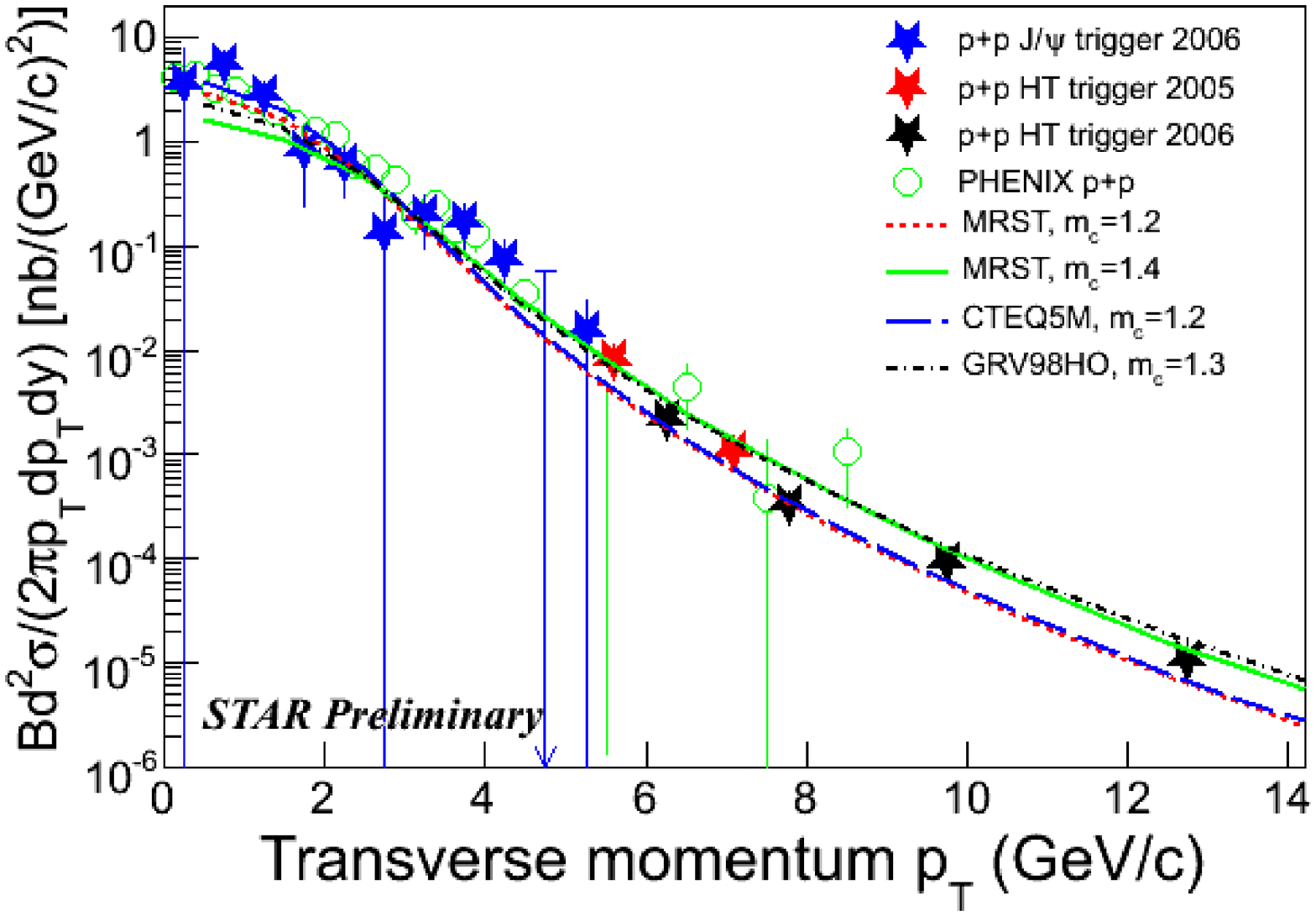}}
      }
      {
      \label{fig:jpsiPt2}
      \includegraphics[width=0.48\textwidth]{{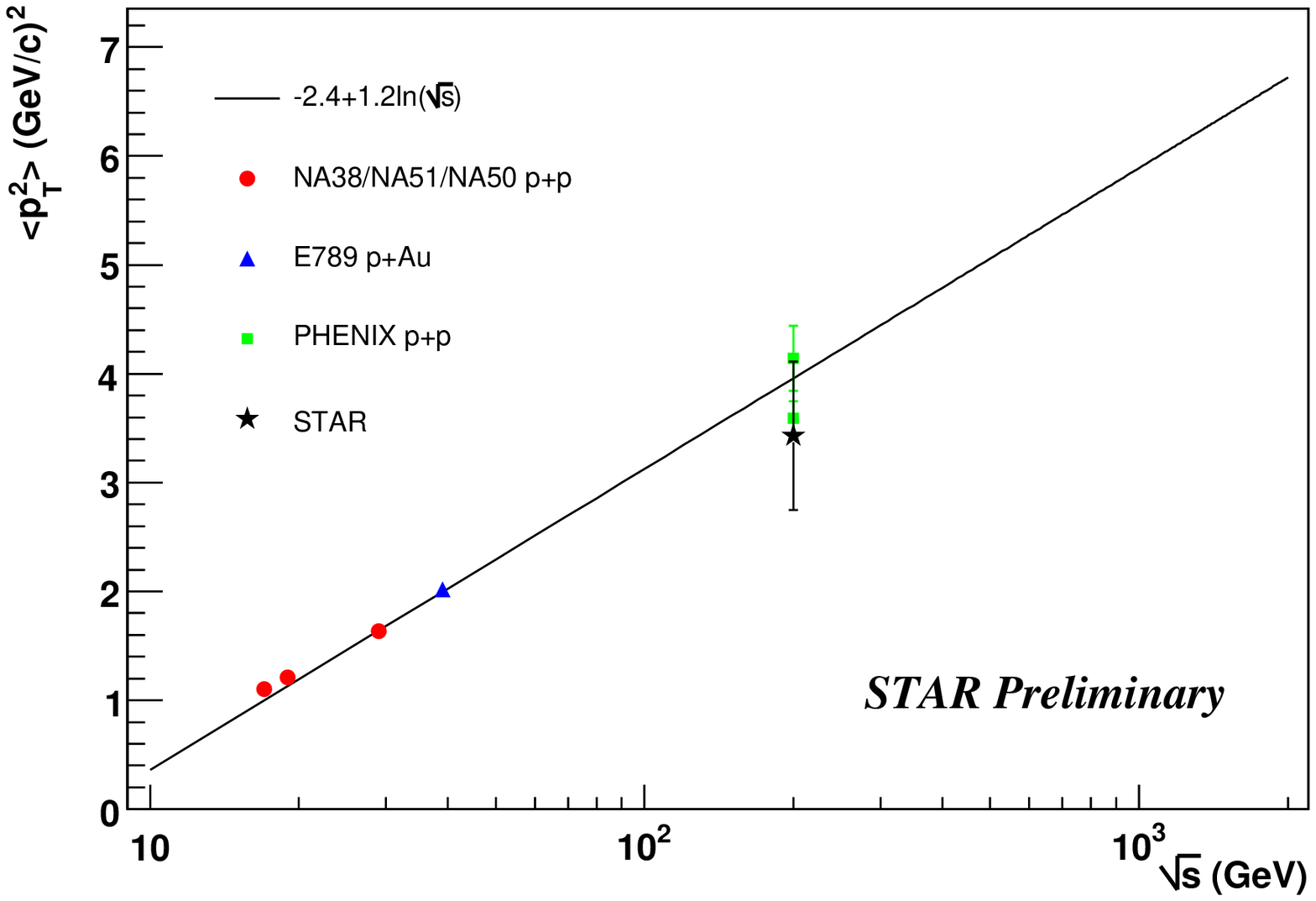}}
      }
      {
      \label{fig:jpsiRaa}
      \includegraphics[width=0.45\textwidth]{{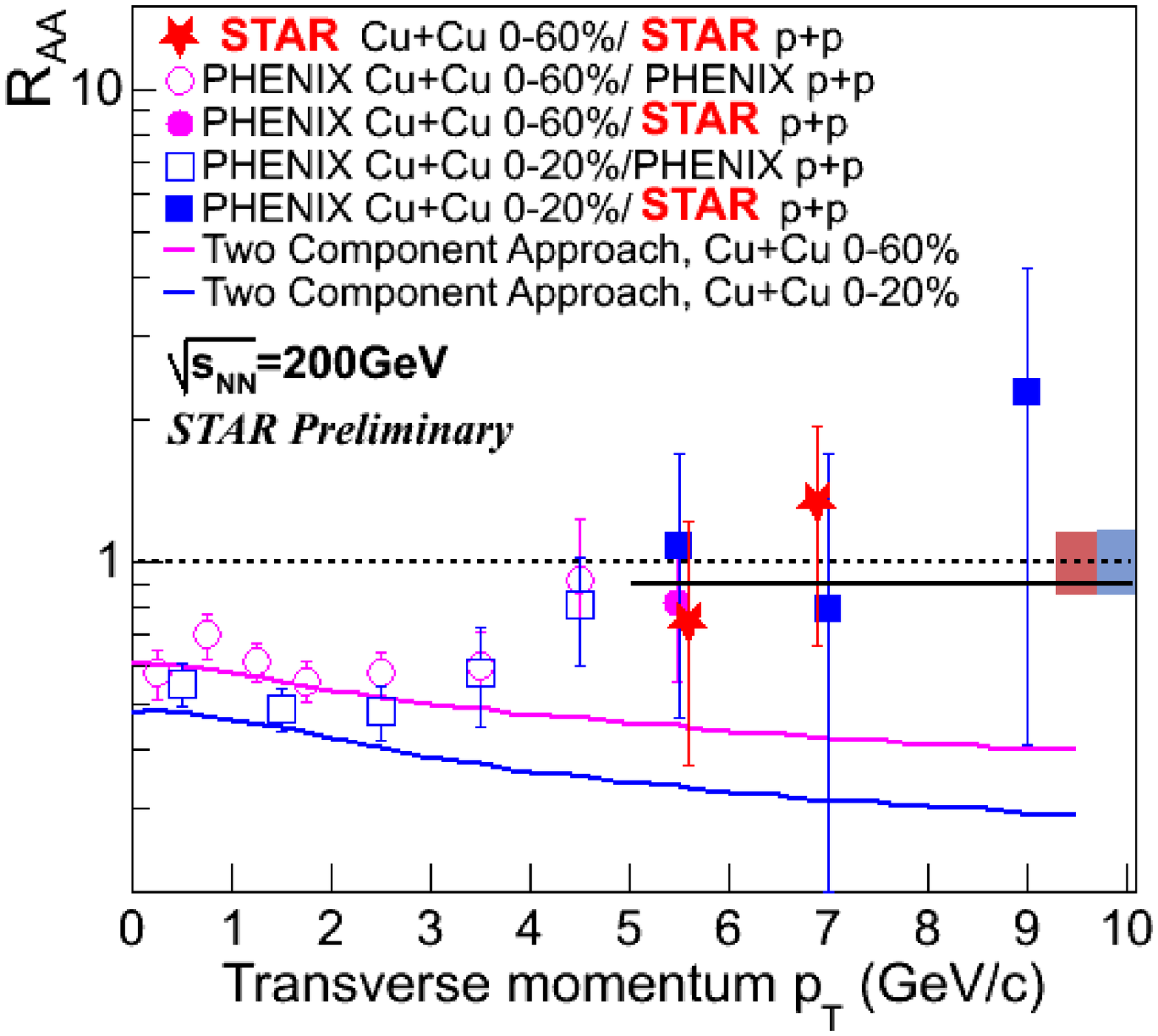}}
      }
    \end{center}
    \caption{$J/\psi$ invariant mass spectrum from di-electron pairs and $p_T$ spectrum.~\cite{mrcQM08,zebo}}
    \label{fig:jpsiMass}
    \end{figure}  
 
\section{Analysis and Results}\label{results}
The quarkonium analysis must begin with an efficient and clean electron identification, since the measurements rely on the di-electron decay channel. The STAR electron identification makes use of the TPC $dE/dx\times p$ information, the BEMC towers energy and, in some cases, the BEMC-SMD information. Reference~\cite{upsilon} details the identification for the measurements made with the specific trigger.
For the high-$p_T$ $J/\psi$ there are extra SMD cuts in the electron identification. These cuts require at least 2 hits in both SMD planes, and also require the position of the cluster to be $\vert\phi_{tow}-\phi_{SMD}\vert<$0.01 and $\vert z_{tow}-z_{SMD}\vert<$2.

Once the identification is done, the di-electron pairs have their invariant mass reconstructed. In the high-$p_T$ $J/\psi$ analysis~\cite{zebo} the reconstruction is made taking one high-$p_T$ electron, identified with TPC+BEMC (tower+SMD) and combining it with a low momentum electron identified only by TPC information.

  \begin{figure}
    \begin{center}
      {
      \label{fig:upsRapPP}
      \includegraphics[width=0.48\textwidth]{{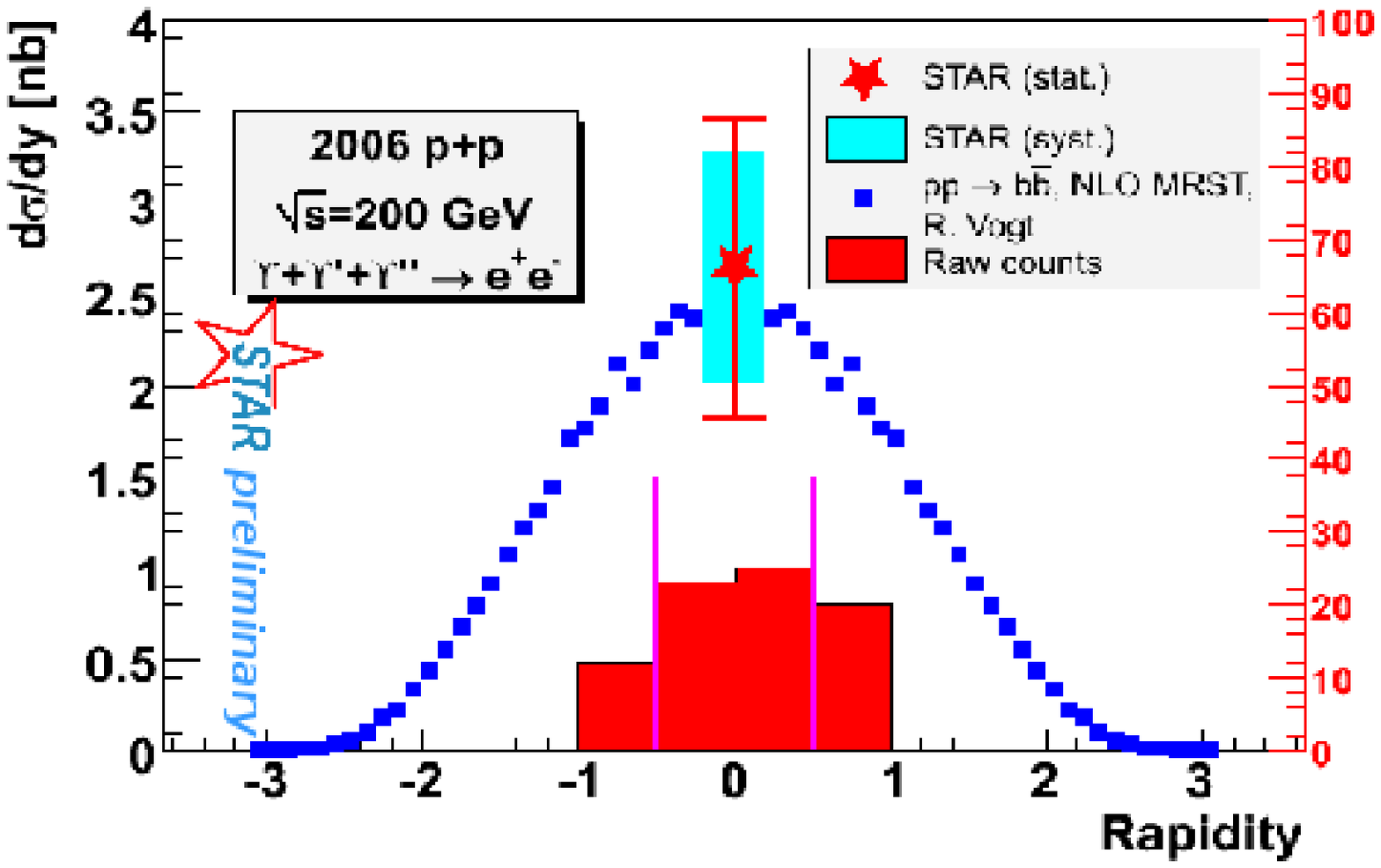}}
      }
      {
      \label{fig:upsEnergPP}
      \includegraphics[width=0.48\textwidth]{{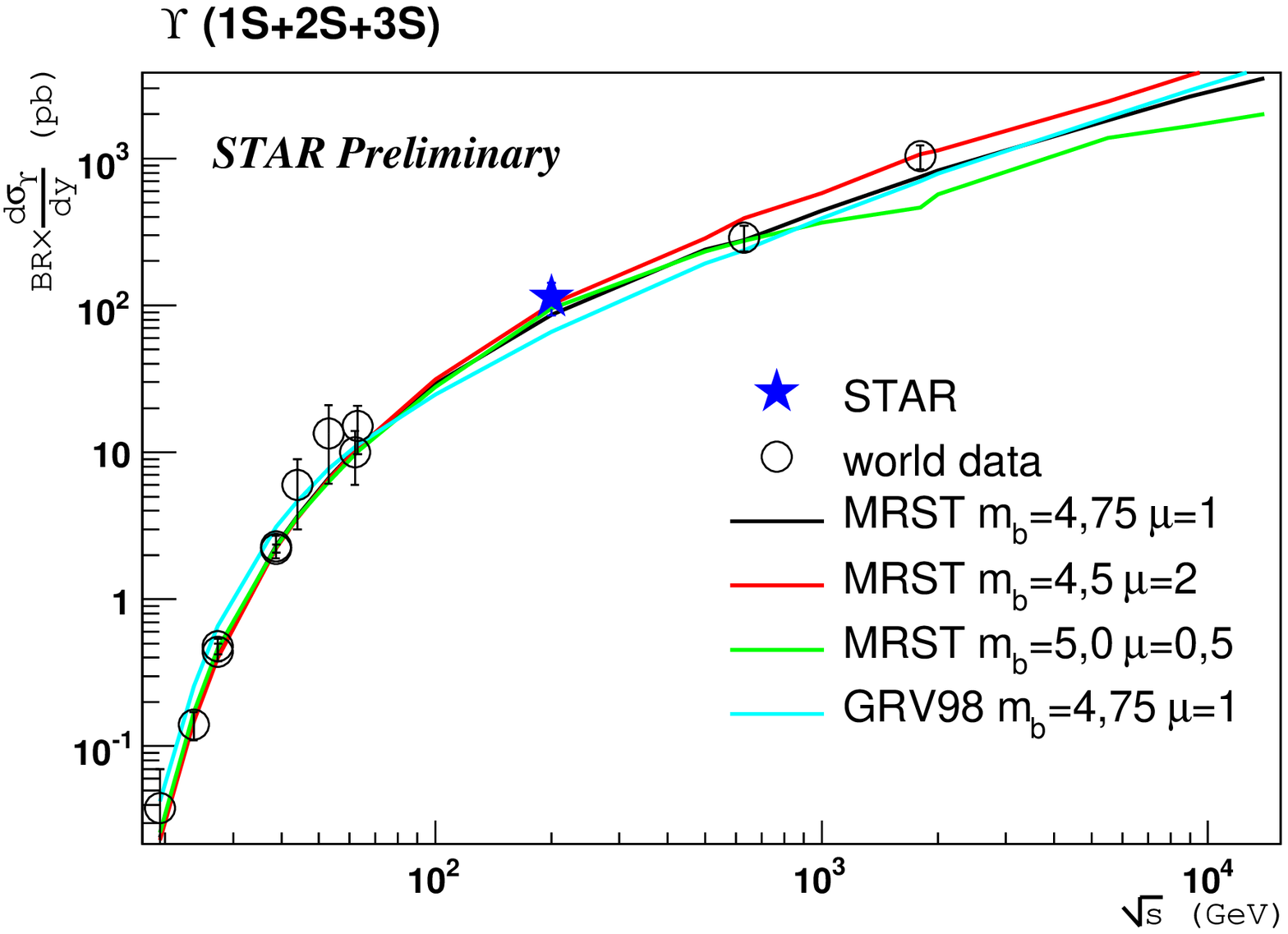}}
      }
      {
      \label{fig:upsSBGAuAu}
      \includegraphics[width=0.44\textwidth]{{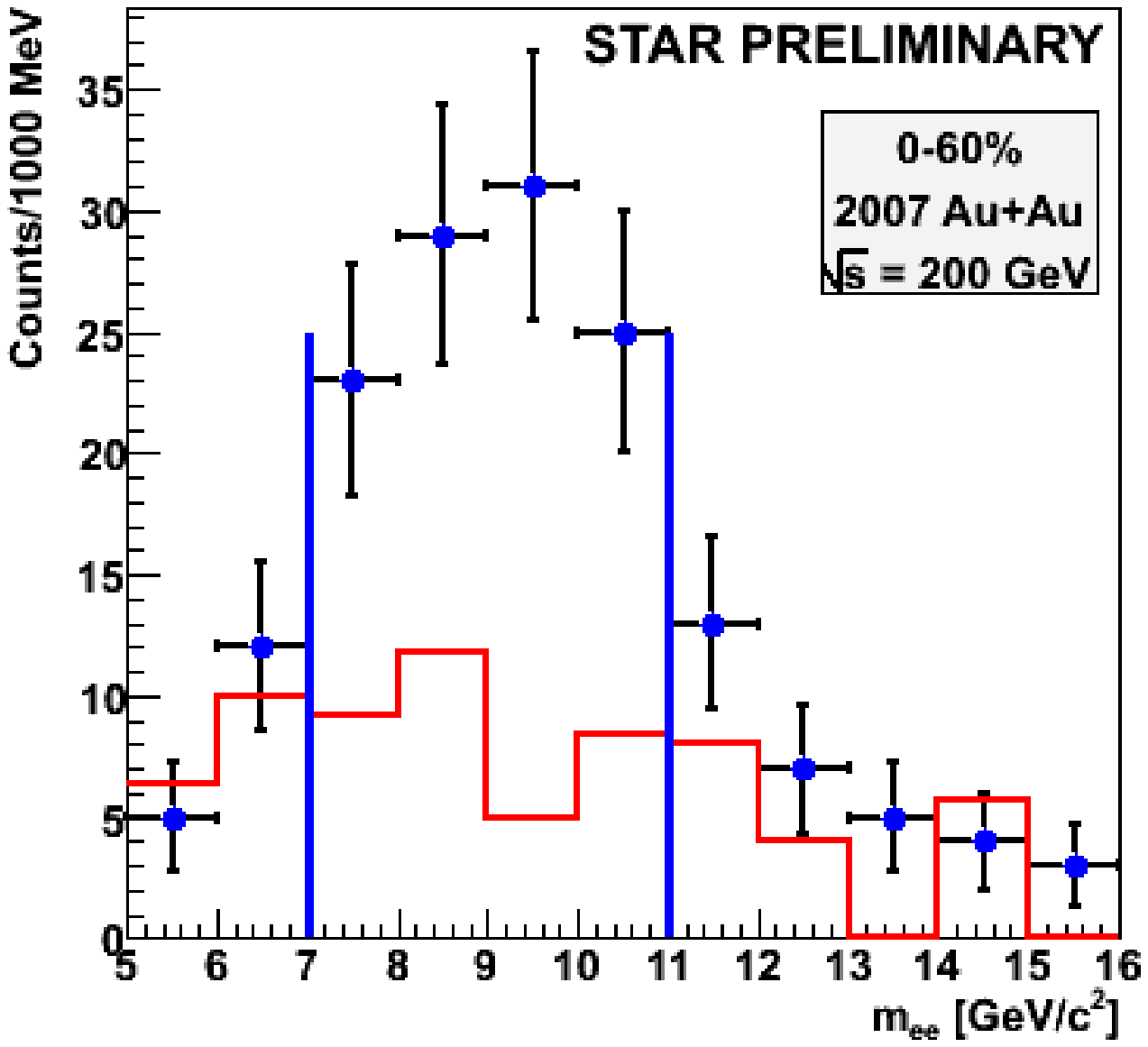}}
      }
      {
      \label{fig:upsSigAuAu}
      \includegraphics[width=0.44\textwidth]{{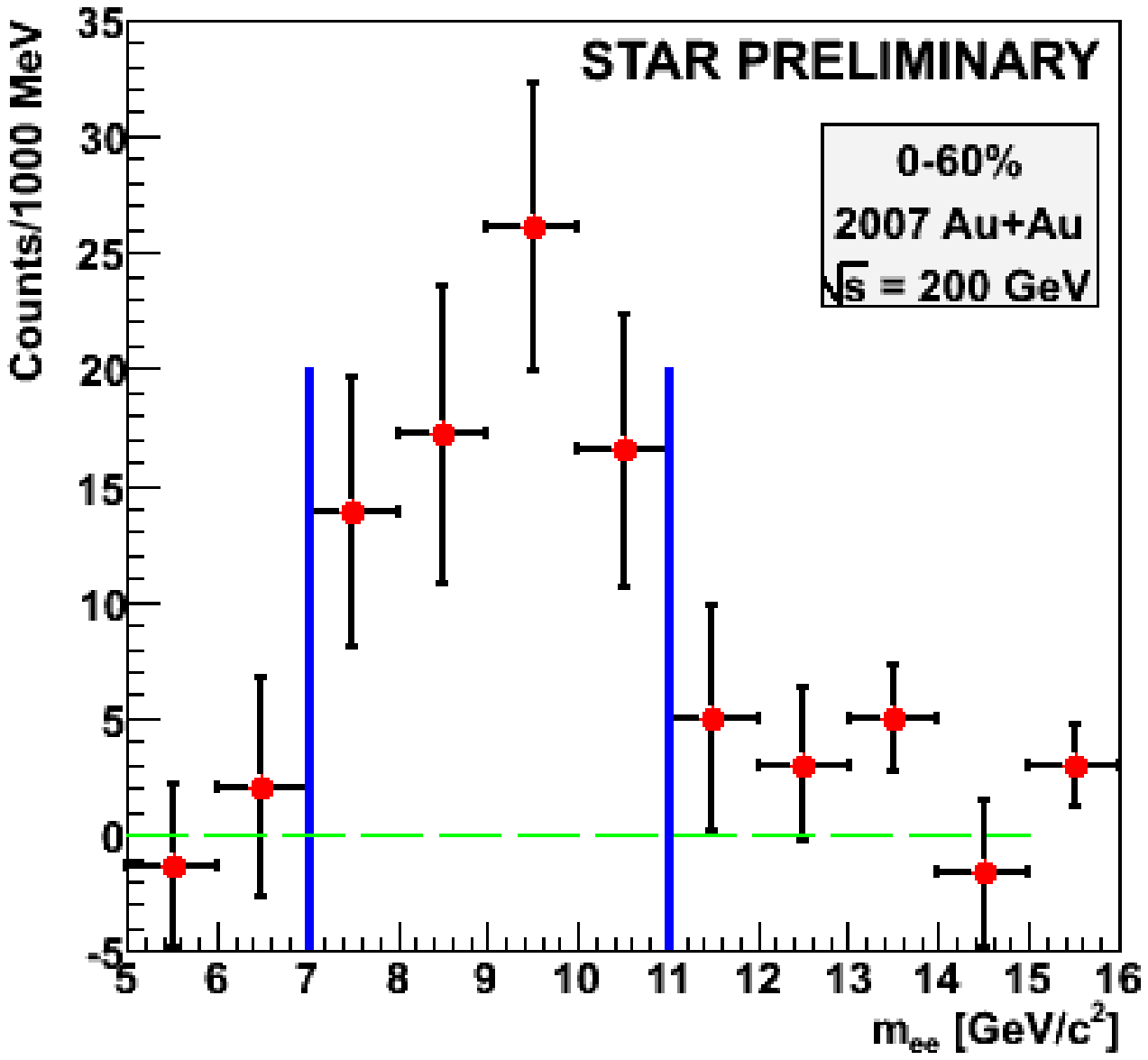}}
      }
    \end{center}
    \caption{$\Upsilon$ results in p+p \cite{upsilon} and Au+Au \cite{debasish}.}
    \label{fig:ups}
    \end{figure}  
\subsection{$J/\psi$ results}\label{sub:jpsi}
The reconstructed mass\footnote{The high-$p_T$ p+p mass spectra, and Cu+Cu data are omitted to save space, but can be found in \cite{zebo}.} and $p_T$ spectra for $J/\psi$ in p+p are presented in figure \ref{fig:jpsiMass}. In the upper left the invariant mass subtracted from the combinatorial background with a GEANT~\cite{Geant3} simulation fitted to the data. The only fitted parameter is the height of the $J/\psi$ peak, showing that the simulation have a good consistency with our data in the postion and width of the peak. Even the bremsstrahlung tale on the left is present in both, data and simulation. In the upper right is the full $p_T$ spectrum, including both, high-$p_T$ and triggered $J/\psi$'s. The circles are PHENIX data points~\cite{jpsiPhenix} and the lines correspond to pQCD-CEM calculations~\cite{ramona}, where is possible to see the good consistency between our data and PHENIX, and the theoretical calculations describe our data fairly well even at higher $p_T$. Our measurement extends considerably the measured $p_T$ range at this energy. The lower line of the figure presents, at the left, the $\langle p_T^2\rangle$, calculated based on low-$p_T$ data, compared with world data systematics~\cite{wdata}, showing consistency once more with PHENIX and with other experimental data trend. On the right side, the extended $R_{AA}$ for the Cu+Cu measurement show an increase for higher $p_T$ data points, contradicting the model predictions~\cite{hotWind,twoComponent}.

\subsection{$\Upsilon$ results}\label{sub:ups}
In figure \ref{fig:ups} we present the $\Upsilon$ results for p+p and Au+Au collisions at $\sqrt{s}=$200 GeV. The upper line presents p+p results \cite{upsilon} compared with pQCD calculations of the rapidity distribution \cite{ramona} (left) and, in the right panel, the $\sqrt{s}$ dependence of $\frac{d\sigma^{\Upsilon}}{dy}$~\cite{bejidian} as well as world data~\cite{wdata}. Our measurements show consistency with the other experimental data trend in $\sqrt{s}$.

In the second line of figure \ref{fig:ups} we present the first $\Upsilon$ measurement in A+A collisions \cite{debasish}. In this very preliminary stage we present only the invariant mass spectra of signal and combinatorial background.

\section{Conclusions}\label{concl}
In this work we presented the preliminary results of the STAR quarkonium program with exciting measurements of $J/\psi$ and $\Upsilon$ in various collisional systems. The $J/\psi$ $R_{AA}=$0.9$\pm$0.2 at higher $p_T$ challenges the theoretical models. On the $\Upsilon$ side we presented the first measurement in p+p collisions at mid-rapidity and the first measurement in heavy-ion collisions ever! 

The perspectives are the $\Upsilon$ $R_{AA}$ in the short term, and some mid-term upgrades in STAR and in RHIC will enhance the capabilities of quarkonium measurements.
 
\section*{Acknowledgments}
This work was supported by Brookhaven National Laboratory and the Brazilian support agencies CNPq and CAPES.
 
\section*{References}
\bibliography{cosentino_WWND08}

\vfill\eject
\end{document}